\documentclass[manuscript]{emulateapj}
\usepackage{natbib}

\begin{document}

\title{ The ACS Fornax Cluster Survey. XII. Diffuse Star Clusters in Early-type Galaxies }

\author{ Yiqing Liu\altaffilmark{1,2}, Eric W. Peng\altaffilmark{1,2}, Sungsoon Lim\altaffilmark{1,2}, Andr\'es Jord\'an\altaffilmark{3}, John Blakeslee\altaffilmark{4}, Patrick C{\^o}t{\'e}\altaffilmark{4}, Laura Ferrarese\altaffilmark{4}, Petchara Pattarakijwanich\altaffilmark{2,1} }

\altaffiltext{1}{Department of Astronomy, Peking University, Beijing 100871, China; yiqing.liu@pku.edu.cn, peng@pku.edu.cn}
\altaffiltext{2}{Kavli Institute for Astronomy and Astrophysics, Peking University, Beijing 100871, China}
\altaffiltext{3}{Instituto de Astrof\'isica, Facultad de F\'isica, Pontificia Universidad Cat\'olica de Chile, Av. Vicu\~{n}a Mackenna 4860, 7820436 Macul, Santiago, Chile}
\altaffiltext{4}{Herzberg Institute of Astrophysics, National Research Council of Canada, Victoria, BC V9E 2E7, Canada}


\begin{abstract}

Diffuse star clusters (DSCs) are old and dynamically hot stellar systems that have lower surface brightness and more extended morphology than globular clusters (GCs). Using the images from HST/ACS Fornax Cluster Survey, we find that 12 out of 43 early-type galaxies (ETGs) in the Fornax cluster host significant numbers of DSCs. Together with literature data from the HST/ACS Virgo Cluster Survey, where 18 out of 100 ETGs were found to host DSCs, we systematically study the relationship of DSCs with GCs, and their host galaxy environment. Two DSC hosts are post-merger galaxies, with most of the other hosts either having low mass or showing clear disk components. 
We find that while the number ratio of DSCs to GCs is nearly constant in massive galaxies, the DSC-to-GC ratio becomes systematically higher in lower mass hosts. This suggests that DSCs may be more efficient at forming (or surviving) in low density environments. 
DSC hosts are not special either in their position in the cluster, or in the galactic color-magnitude diagram. 
Why some disk and low-mass galaxies host DSCs while others do not is still a puzzle, however. The mean ages of DSC hosts and non-hosts are similar at similar masses, implying that formation efficiency, rather than survival, is the reason behind different DSC number fractions in early-type galaxies. 

\end{abstract}



\section{Introduction}

Globular clusters (GCs) are relatively more massive and compact compared to other kinds of star clustersr. \citet{Misgeld_Hilker_11} show that the surface density of GCs are well correlated with their masses, with the more massive GCs having higher surface densities, and the effective radii are distributed tightly around 3~pc. 

However, this view has been updated with the improvement of our detection ability. Using the Hubble Space Telescope (HST), \citet{Larsen_Brodie_00} discovered a population of old star clusters that have GC-like luminosity but much larger sizes in a nearby S0 galaxy NGC~1023. 
Comparing with GCs, they are redder and mostly fainter than $M_V=-7$ with half-light radii ($r_h$) in the range of 7-15~pc, while the common GCs have a luminosity function peaked at $M_V=-7.4$ and a typical $r_h$ of 3~pc. On the other hand, they are significantly brighter and larger than the open clusters in the Milky Way. 

This discovery opened a new field, rapidly leading to more detections in other galaxies. 
Similar diffuse star clusters (DSCs) were detected in the nearby field galaxies NGC~3384, NGC~5195, NGC~5194 (M51), and NGC~6822 \citep{Larsen_01,Lee_05,Hwang_Lee_08,Hwang_11}, as well as 12 Early Type Galaxies (ETGs) in the Virgo Cluster \citep[hereafter P06]{P06}. They are also detected in the outer halo of our Milky Way and M31, our satellite galaxies, and the dwarf elliptical galaxy Scl-dE1 \citep{vdBergh_Mackey_04,Huxor_05,daCosta_09}. These DSCs tend to have larger $r_h$ (20$-$40~pc), but are still smaller and brighter than the ultra-faint galaxies at similar magnitude. 

Nonetheless, there are galaxies with no DSCs detected. This naturally raises questions: Why are DSCs only detected in certain galaxies, instead of others? Do these galaxies have special physical conditions for DSC formation, or for their survival? Does the DSC formation follow the general picture of star cluster formation? 

The last question is the most fundamental one. Besides the typical way of star cluster formation, tidal stripping of galaxies and mergers of cluster complexes are two candidate mechanisms. 
In the former case, although the galactic cores left from stripping always have large sizes, they usually have relatively high surface brightness, which are more similar to ultra-compact dwarf galaxies (UCDs). 
The merger origin \citep{Fellhauer_Kroupa_02,Burkert_05,Bruns_09,Bruns_11} is disfavored. 
\citet{Assmann_11} found that the velocity dispersions of merger-produced DSCs would be too high. Furthermore, they also excluded the scenario in which DSCs formed by expanding normal star clusters due to the gas expulsion or stellar mass loss during their early evolution, as the observed star formation efficiency is not high enough. 

Therefore, DSCs probably form in a way similar to other star clusters. Then the question remains as to why they only exist in certain galaxies. 
Possibly, DSC formation may require special environmental conditions. \citet{Harris_Pudritz_94} argue that the supergiant molecular clouds that form massive star clusters are pressure-confined by the interstellar medium (ISM) of their parent galaxies. Furthermore, \citet{Mclaughlin_00} noted a relation between the binding energy and the galactocentric distance of the Milky Way globular clusters. These all imply that the more extended star clusters prefer to form in lower density regions. A more directed study is from \citet{Elmegreen_08}, who suggested that the difference between star formation in bound clusters and in loose groupings is attributed to the difference in cloud pressure. High-pressure regions place higher fraction of stars in bound clusters, while low-pressure regions prefer to make unbound stellar groupings; and the regions with moderately low density and moderately high Mach number would produce low-density bound clusters like DSCs. 

Low-mass galaxies provide such environments. Based on the evidence that extended star clusters are found in dwarf galaxies NGC~6822 and Scl-dE1, the low-mass halo origin is plausible, and the DSCs which are observed in the outer halo of massive galaxies can be explained by accretion from low-mass satellite galaxies. 
Moreover, \citet{Masters_10} showed a trend between GC size and host galaxy mass, with the fainter galaxies have larger GCs. 
Galactic disks are another such low density environment. For example, \citet{Pellerin_10} suggested that in a collisional ring galaxy NGC~922, the highly shocked low density ring which contains a number of star forming complexes and young massive clusters is a possible place for forming DSCs. Among all the previously founded DSC host galaxies, most are either dwarf or disky galaxies. In addition, DSCs in NGC~1023 have systematic rotation curve similar to the host galaxy \citep{Larsen_Brodie_02}. 

However, not all low-mass or disk galaxies are associated with DSCs. So the question that naturally follows is: are those DSC host galaxies different from their counterparts, or they are just in a stage of evolution when DSCs have not been entirely disrupted? 
Using N-body simulations, \citet{Hurley_Mackey_10} found that DSCs can form naturally within weak tidal fields, which provides a possible scenario that the detected DSCs are just the ones that have not been tidally disrupted, because the disruption timescale is small when star clusters have larger radii \citep{Gnedin_99}. 

To further investigate these questions, a large and complete sample is necessary. Because of their low luminosities, DSC studies are limited to the nearby universe, and the sample from the literature is not big because the frequency of their appearance is relatively low. Moreover, except for the Virgo Cluster, which is the nearest galaxy cluster (16.5~Mpc away) that has been examined by P06, no other cluster environment has been used for DSC studies. Therefore, in order to build a larger sample for DSC study, we turn to the Fornax cluster, which is the second nearest cluster located 20~Mpc away. 

Space-based imaging is a powerful technique to detect these small low surface brightness DSCs. Previous work by P06 used the data from the ACS Virgo Cluster Survey (ACSVCS; \citealt{Cote_04}) to study the DSCs in that cluster. This work uses the data from the ACS Fornax Cluster Survey (ACSFCS; \citealt{Jordan_07}), which is a complementary program to the ACSVCS that imaged 43 galaxies in the Fornax cluster, to perform similar studies.
We compare DSCs and GCs using this larger sample, and look for their dependence on galactic properties such as type, mass, and environment. A special advantage of this work is that these two surveys have the same instrument setups and data reduction processes, which aids in our comparison. 

The paper is structured as follow: 
Our data are introduced in $\S$\ref{obs}. The selection and basic properties of DSCs are described in $\S$\ref{selection}. Then we investigate the properties of DSC host galaxies in $\S$\ref{galaxy}, and compare the color, spacial distribution, and formation efficiency of DSCs and GCs in $\S$\ref{D_G}. Possible DSC formation scenarios are discussed in $\S$\ref{discussion}. Conclusions are summarized in $\S$\ref{conclusion}. \\


\section{Observations} 
\label{obs}

The ACSFCS \citep{Jordan_07} is a program that has imaged 43 ETGs in the Fornax Cluster with the HST/ACS. This is a complete sample of Fornax galaxies brighter than $B_T\sim15.5$ ($M_B\sim-16$) mag, covering the morphological types of E, S0, SB0, dE, dE,N, dS0, or dS0,N. It includes 41 galaxies from the Fornax Cluster Catalog (FCC; \citealt{Ferguson_89a}), as well as 2 outlying elliptical galaxies NGC~1340 and IC~2006. This survey took $202\arcsec \times 202\arcsec$ field of view (FOV) images for each galaxy in F475W and F850LP filters, with a pixel scale of $0.049\arcsec$. These two filters are roughly the same as the SDSS $g$ and $z$ bands (hereafter referred to as $g$ and $z$ band), and they are sensitive to metallicity and age of stellar populations. Because a primary science goal of the program is to study extragalactic globular clusters, the images are sufficiently deep that $\sim90\%$ of the GCs can be detected at a high level of completeness \citep{Cote_04} with a high spatial resolution. Moreover, the contaminants of background galaxies have been simulated by using 16 blank high-latitude control field images from HST archive, as in P06. 

We also use data from the ACSVCS, with the identical instrument setup. 
The ACSVCS sample contains 100 ETGs with $B_T<16$, but is only complete to $B_T<12.15$ ($M_B<-18.94$). In the luminosity range where the sample is incomplete, 63 low-mass galaxies were removed from the sample. 
The data reduction of both surveys was performed in the same way, following P06.
One exception is on the star cluster candidates larger than 10~pc. For these objects from ACSVCS, their structural parameters were measured precisely by preforming a new model of profile fitting. However, it was not applied for ACSFCS, and we limit our sample to the objects smaller than 10~pc in this study. \\


\section{DSC Selection}
\label{selection}

The data reduction process is described in \citet{Jordan_04}, for both image analysis and point source selection. Among the output of GC candidates, \citet{Jordan_09} evaluated the probability $p_{GC}$ that a given object is a GC, according to its position in the size-magnitude parameter space. All the basic parameters of the GC candidates from ACSFCS are listed in \citet{Jordan_15}. 
In previous ACSVCS and ACSFCS studies, $p_{GC} \ge 0.5$ is used to select GCs, and we use the same criterion in this work. Usually, those objects with $p_{GC} < 0.5$ are not as concentrated as GCs and mainly consisted of background galaxies. However, since the expected number of background contaminants has been estimated from control fields, if the number of diffuse objects in a galaxy field significantly exceeds the expectation, we can infer that this galaxy hosts some DSCs. 

Following P06, we select those extended, background-liked DSCs using the criteria $p_{GC} \le 0.2$ and projected half-light radius $r_h \ge 4$~pc (typical GCs have median $r_h\sim3$~pc), avoiding most traditional GCs. 
This selection would leave a fraction of star clusters that are classified into neither GCs nor DSCs, but it is reasonable in this study. Because our primary goal is making a sample of star clusters that are significantly more diffuse than traditional GCs, instead of counting their absolute numbers. 

\begin{figure}
\epsscale{1.3}
\plotone{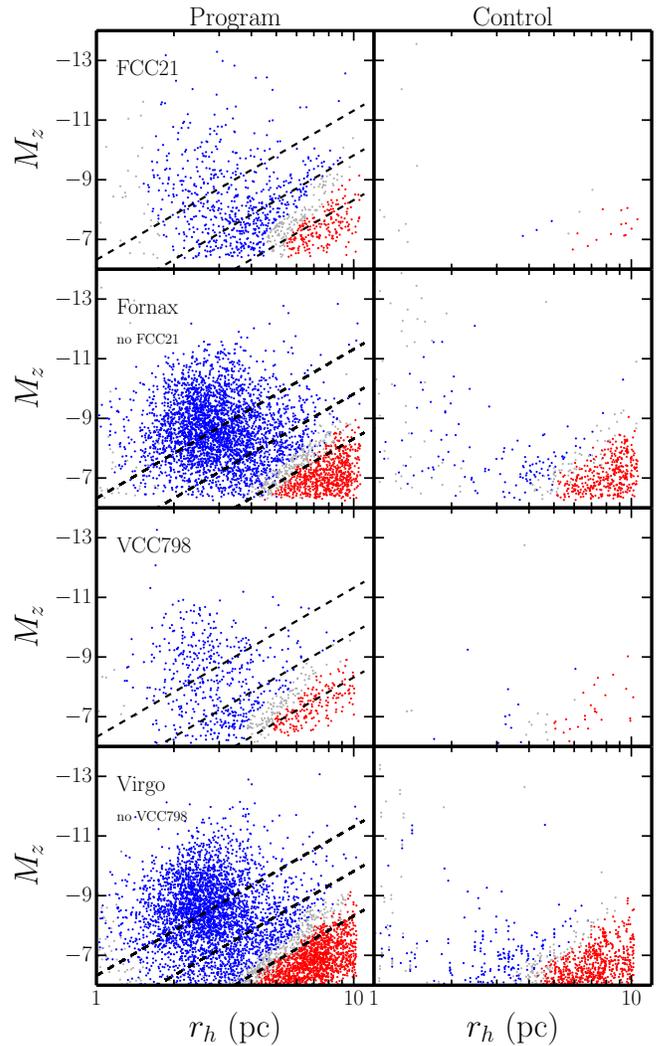}
\caption{ From \textit{top} to \textit{bottom}: Size-magnitude diagrams of all the star cluster candidates with $p_{GC} > 0$ from FCC~21, the combination of the rest of Fornax galaxies with DSC excess, VCC~798, and the combination of the rest of Virgo galaxies with DSC excess. Blue, red and grey points represent GC, DSC, and the rest star cluster candidates respectively. 
The \textit{left} and \textit{right} columns are for program and a randomly selected control fields respectively, and the number of DSC candidates selected from the program field is clearly excess the contaminations from control field.
Diagonal dash lines show the constant mean surface brightness of 18.0, 19.5, 21.0~mag/arcsec$^2$ in $z$-band. Our selection criteria are good at separating DSCs and GCs by surface brightness. \\
  \label{rhMz} }
\end{figure}
\begin{figure}
\epsscale{1.25}
\plotone{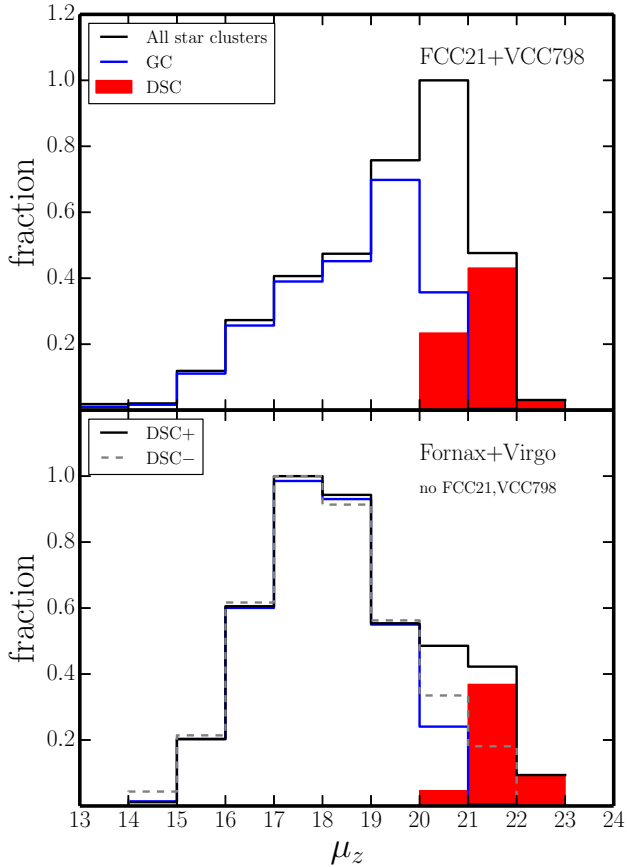}
\caption{ The surface brightness distributions of the star clusters in our sample. The upper panel shows the distributions in the two merger remnants with the most massive DSC systems, and the bottom shows those in the rest galaxies. In both panels, the black, blue, and red histograms represent the distributions of the entire star cluster systems, GCs, and DSCs in DSC host galaxies that normalized by the bin with the highest number of all star clusters respectively. In the bottom panel, the grey dash line shows the distribution of all star cluster candidates from the DSC non-excess galaxies, and normalized by the highest bin. The background contaminants are subtracted in each bin. 
The distribution of star clusters in merger remnants has a peak too faint to tell substantial information about DSCs. However, in the normal ETGs, the distribution of star cluster candidates in the DSC host galaxies has faint excess comparing to that of the DSC non-hosts, which is mainly contributed by the DSCs. It indicates that DSCs are essentially a distinct population of star clusters. \\
  \label{histSB} }
\end{figure}

Figure~\ref{rhMz} shows our selection in the parameter spaces. All the GC candidates with $p_{GC} > 0$ from the DSC-excess galaxies (13 from Fornax and 19 from Virgo, which will be described below) are displayed in the $r_h$-$M_z$ diagrams. The left and right columns are for program and randomly selected control fields respectively. 
From top to bottom, the samples are from the star cluster systems of FCC~21 (NGC~1316; Fornax~A), the combination of the rest of Fornax galaxies with DSC excess, VCC~798 (NGC~4382; M85), and the combination of the rest of Virgo galaxies with DSC excess. 
FCC~21 and VCC~798 have the highest number of DSCs in Fornax and Virgo respectively (Figure~\ref{findDSC} and Table~\ref{gal_DSC} in both this paper and P06). We show these galaxies separately to show the distribution clearly, especially because they might dominate the total distribution by large numbers. The candidates that agreed with the criteria of GC and DSC are shown in blue and red respectively, and the rest are plotted in grey. 
The constant $z$-band mean surface brightness ($\mu_z$) of 18.0, 19.5, 21.0~mag/arcsec$^2$ are marked by diagonal dash lines. 
From these diagrams, all the star clusters distributed continuously in $r_h$-$M_z$ space, and our criteria are as good at separating them as using surface brightness in all host galaxies. 
The DSC candidates are located at the faint end of GC luminosity distributions, but this may be just a selection effect, because we select DSCs with faint surface brightness in a limited range of sizes. The parameter that fundamentally makes DSCs special is the surface brightness $\mu$, which is a combination of luminosity and size. 

Figure~\ref{histSB} displays the surface brightness distributions of star clusters in our sample. We divide the DSC host galaxies into two groups. One consists of the two merger remnants with the most massive DSC systems, FCC~21 and VCC~798 (the upper panel), and the other is made up by the rest galaxies (the bottom panel). In both panels, the black, blue, and red histograms represent the distributions of the entire star cluster systems, GCs, and DSCs in DSC host galaxies that normalized by the bin with the highest number of all star clusters respectively. In the bottom panel, the grey dash line shows the distribution of all star cluster candidates from the DSC non-excess galaxies, and normalized by the highest bin. The background contaminants are subtracted in each bin. 
In the merger remnants, the surface brightness distribution of their all star cluster candidates possibly peaks at a magnitude fainter than our detection limit. Because DSCs occupy the faint end of this distribution, it is hard to infer their substantial behavior in this work. 
From the bottom panel, the distribution of the star clusters in DSC host galaxies is more extended than that of the DSC non-hosts at the fainter end, while they are similar at the bright end and have peaks at similar magnitude. In addition, the distribution of the DSC non-host is symmetric and the faint excess of DSC host galaxies is mainly contributed by the DSC candidates. It indicates that DSCs are essentially a distinct population of star clusters. 

\begin{figure}
\plotone{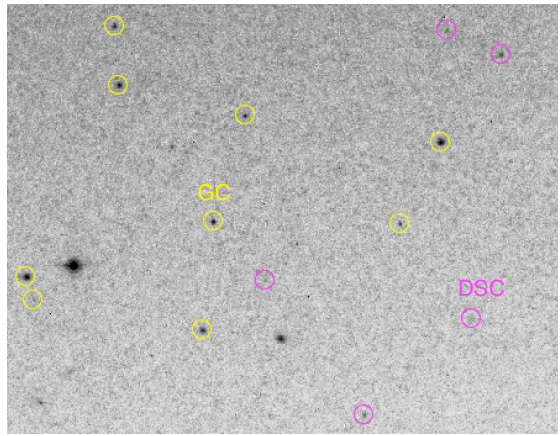}
\caption{Real image of a program field to give an intuitive sense. DSC and GC candidates are shown in magenta and yellow circles respectively. 
\label{im_zoom} }
\end{figure}

\begin{figure}
\epsscale{1.2}
\plotone{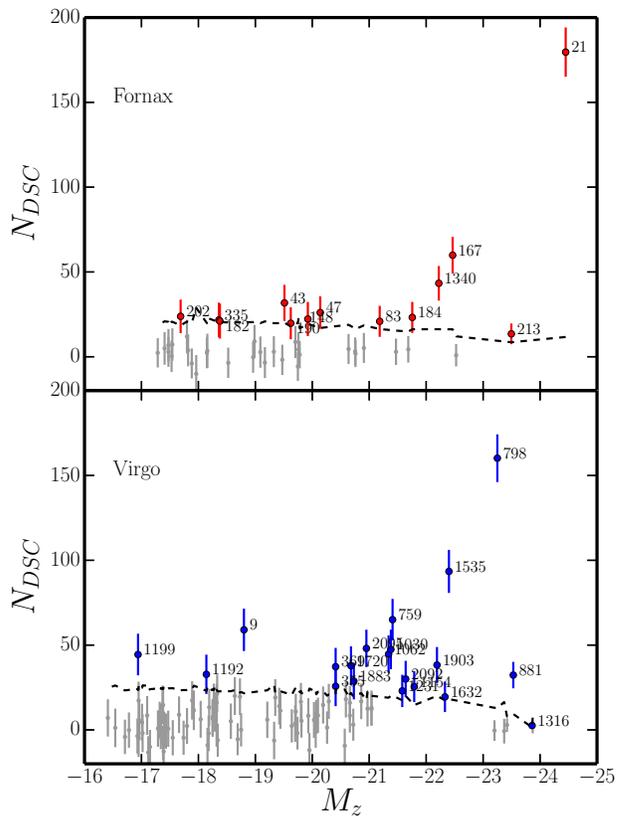}
\caption{ 
The number of DSCs detected in program fields with 1$\sigma$ error bars of the ACSFCS (\textit{upper}) and ACSVCS (\textit{bottom}) sample. Thirteen ETGs in Fornax and twenty ETGs in Virgo with DSC numbers that 3$\sigma$ higher than the contamination level are displayed in red and blue circles respectively. The dash lines show the 3$\sigma$ level of the control fields. The FCC and VCC IDs of the host galaxies are written next to their data points. \\
  \label{findDSC} }
\end{figure}

Figure~\ref{im_zoom} shows how GC and DSC candidates look like on the image. GC and DSC candidates are circled in yellow and magenta respectively. DSCs are less compact than GCs, and some are not well separated from background galaxies. 

\begin{deluxetable*}{llcccccl}
\tabletypesize{\scriptsize}
\tablecaption{Properties of the galaxies with DSC excess in our sample. \label{gal_DSC}}
\tablewidth{0pt}
\tablehead{
\colhead{FCC} & \colhead{NGC} &
\colhead{RA (h m s)} & \colhead{Dec (d m s)} &
\colhead{$M_z$} & \colhead{g$-$z} & 
\colhead{$N_{DSC}$} & 
\colhead{Type} \\
\colhead{} & \colhead{} & 
\colhead{(J2000)} & \colhead{(J2000)} &
\colhead{(mag)} & \colhead{(mag)} & 
\colhead{} & 
\colhead{}
}
\startdata
 21 & 1316 & $03:22:42.09$ & $-37:12:31.63$ & $-24.45$ & $1.37$ & $179.6\pm 14.5$ & S0(pec)\\
 213 & 1399 & $03:38:29.14$ & $-35:27:02.30$ & $-23.50$ & $1.41$ & $13.5 \pm 6.1 $ & E0 \\
 -- & 1340 & $03:28:19.70$ & $-31:04:05.00$ & $-22.22$ & $1.33$ & $43.3 \pm 10.2$ & E5 \\
 167 & 1380 & $03:36:27.45$ & $-34:58:31.09$ & $-22.46$ & $1.31$ & $59.9 \pm 10.8$ & S0/a \\
 83 & 1351 & $03:30:35.04$ & $-34:51:14.51$ & $-21.18$ & $1.39$ & $20.9 \pm 9.2 $ & E5 \\
 184 & 1387 & $03:36:56.84$ & $-35:30:23.85$ & $-21.75$ & $1.59$ & $23.2 \pm 9.2 $ & SB0 \\
 47 & 1336 & $03:26:31.97$ & $-35:42:44.59$ & $-20.14$ & $1.28$ & $26.1 \pm 9.6 $ & E4 \\
 43 & IC~1919 & $03:26:02.30$ & $-32:53:36.80$ & $-19.51$ & $1.15$ & $31.7 \pm 10.7$ & dS0/2(5),N \\
 190 & 1380B & $03:37:08.86$ & $-35:11:37.54$ & $-19.62$ & $1.37$ & $19.7 \pm 9.4 $ & SB0 \\
 148 & 1375 & $03:35:16.79$ & $-35:15:55.95$ & $-19.92$ & $1.21$ & $22.2 \pm 10.1$ & S0(cross) \\
 335 & -- & $03:50:36.64$ & $-35:54:29.27$ & $-18.36$ & $1.13$ & $21.7 \pm 10.2$ & E \\
 182 & -- & $03:36:54.24$ & $-35:22:22.69$ & $-18.38$ & $1.34$ & $21.0 \pm 10.2$ & S0 pec \\
 202 & 1396 & $03:38:06.40$ & $-35:26:17.96$ & $-17.69$ & $1.19$ & $23.8 \pm 9.9 $ & dE6,N \\
\enddata
\tablecomments{ Coordinates are from \citet{Jordan_04}. $M_z$ and $g-z$ are derived from model fits to the HST/ACS images of these galaxies and the distance modules in \citet{Blakeslee_09}. $g-z$ is from \citet{Blakeslee_09}. $N_{DSC}$ is the number of DSCs selected from program images subtracted by the mean number of contaminants from 16 control fields, with the 1$\sigma$ uncertainty that estimated from Poisson distributions. The galaxy classifications are from \citet{Ferguson_89a}. } 
\end{deluxetable*}
\begin{deluxetable*}{llcccccl}
\tabletypesize{\scriptsize}
\tablecaption{Properties the galaxies with DSC excess in the ACSVCS sample. \label{galV_DSC}}
\tablewidth{0pt}
\tablehead{
\colhead{VCC} & \colhead{NGC} &
\colhead{RA (h m s)} & \colhead{Dec (d m s)} &
\colhead{$M_z$} & \colhead{g$-$z} & 
\colhead{$N_{DSC}$} & 
\colhead{Type} \\
\colhead{} & \colhead{} & 
\colhead{(J2000)} & \colhead{(J2000)} &
\colhead{(mag)} & \colhead{(mag)} & 
\colhead{} & 
\colhead{}
}
\startdata
881  & 4406    & $12:26:11.74$ & $+12:56:46.4$ & $-23.53$ & $1.57$ & $32.3 \pm 7.79 $ & S0$_1$(3)/E3\\
798  & 4382    & $12:25:24.04$ & $+18:11:25.9$ & $-23.25$ & $1.38$ & $160.2\pm 14.05$ & S0$_1$(3)   \\
1535 & 4526    & $12:34:03.10$ & $+07:41:59.0$ & $-22.40$ &  --    & $93.4 \pm 12.71$ & S0$_3$(6)   \\
1903 & 4621    & $12:42:02.40$ & $+11:38:48.0$ & $-22.19$ & $1.53$ & $38.2 \pm 10.62$ & E4         \\
1632 & 4552    & $12:35:39.82$ & $+12:33:22.6$ & $-22.33$ & $1.61$ & $19.5 \pm 8.98 $ & S0$_1$(0)\\
1231 & 4473    & $12:29:48.87$ & $+13:25:45.7$ & $-21.58$ & $1.53$ & $23.0 \pm 9.53 $ & E5         \\
2095 & 4762    & $12:52:56.00$ & $+11:13:53.0$ & $-20.95$ & $1.44$ & $48.0 \pm 11.01$ & S0$_1$(9) \\
1154 & 4459    & $12:29:00.03$ & $+13:58:42.9$ & $-21.79$ & $1.44$ & $25.5 \pm 9.09 $ & S0$_3$(2)\\
1062 & 4442    & $12:28:03.90$ & $+09:48:14.0$ & $-21.34$ & $1.53$ & $44.7 \pm 11.04$ & SB0$_1$(6) \\
2092 & 4754    & $12:52:17.50$ & $+11:18:50.0$ & $-21.64$ & $1.50$ & $30.0 \pm 10.70$ & SB0$_1$(5) \\
369  & 4267    & $12:19:45.42$ & $+12:47:54.3$ & $-20.41$ & $1.57$ & $37.2 \pm 11.08$ & SB0$_1$ \\      
759  & 4371    & $12:24:55.50$ & $+11:42:15.0$ & $-21.41$ & $1.54$ & $65.0 \pm 12.22$ & SB0$_2$(r)(3)\\
1030 & 4435    & $12:27:40.49$ & $+13:04:44.2$ & $-21.38$ &  --    & $47.4 \pm 11.65$ & SB0$_1$(6) \\
1720 & 4578    & $12:37:30.61$ & $+09:33:18.8$ & $-20.68$ & $1.44$ & $37.8 \pm 11.43$ & S0$_{1/2}$(4)\\
355  & 4262    & $12:19:30.61$ & $+14:52:41.4$ & $-20.41$ & $1.52$ & $25.7 \pm 11.68$ & SB0$_{2/3}$ \\
1883 & 4612    & $12:41:32.70$ & $+07:18:53.0$ & $-20.73$ & $1.32$ & $28.7 \pm 10.76$ & RSB0$_{1/2}$\\
9    & IC~3019 & $12:09:22.34$ & $+13:59:33.1$ & $-18.80$ & $1.15$ & $59.0 \pm 12.51$ & dE1,N      \\
1192 & 4467    & $12:29:30.20$ & $+07:59:34.0$ & $-18.14$ & $1.52$ & $32.7 \pm 11.60$ & E3         \\
1199 & IC~3602 & $12:29:34.97$ & $+08:03:31.4$ & $-16.94$ & $1.56$ & $44.4 \pm 12.26$ & E2         \\  
\enddata
\tablecomments{ Coordinates, $M_z$, and $g-z$ are from \citet{Cote_04}, \citet{P08}, and \citet{Ferrarese_06} respectively. $N_{DSC}$ and the 1$\sigma$ uncertainty are derived in the same way as in Table~\ref{gal_DSC}. The galaxy classifications are from \citet{Ferguson_89a}. } 
\end{deluxetable*}

Using such selection criteria, we claim that a galaxy hosts DSCs if the net number of diffuse objects (the number detected in the program field without completeness correction but subtracted by the mean number of contaminants from the 16 control fields) is 3$\sigma$ higher than the mean number of contaminants in the control fields, where the $\sigma$ is the standard deviation of the number counts from the 16 control fields. 
The upper panel of Figure~\ref{findDSC} shows that 13 galaxies in our sample have significant number of DSCs. 
However, because the star cluster system of FCC~202 belongs to the halo of the bright central galaxy (BCG) FCC~213 (NGC~1399), only 12 Fornax galaxies contain DSCs substantially. 
Basic parameters of these galaxies are listed in Table~\ref{gal_DSC}. 
The errors are estimated as the 1$\sigma$ uncertainty from Poisson distributions, and the standard deviations of contaminants from the 16 control fields are considered. 

Because a large fraction of the low-mass host galaxies have ambiguous excess, we preform a test with the criteria of $p_{GC} < 0.5$ and $r_h \ge 7$~pc. This is similar to the cut used for "faint fuzzies" in other works (e.g. \citealt{Larsen_Brodie_00}), which are essentially the same objects as the DSCs we are studying. 
Under the alternative criteria, the same 13 galaxies are selected out, as well as FCC~177, which is at the boundary of the cut. Therefore, we conclude all the 13 galaxies as DSC hosts in this work. 

Our selection of DSC host galaxies is different from P06. P06 also defined DSC hosts as 3$\sigma$ higher than the background, but the $\sigma$ was the errors of the number of DSC measurements instead of the scatter of background contaminants. Therefore, we use the new criterion to select DSC hosts from Virgo in this work, and the bottom panel of Figure~\ref{findDSC} shows that 20 ETGs in Virgo have DSC number excess. 
However, because the BCG VCC~1316 (M87) only have 3 DSC candidates with an expected background of $0.44\pm0.50$, we remove it from the DSC host galaxies. 
Besides, similar to the case of FCC~202 in Fornax, the DSC systems of VCC~1192 and VCC~1199 belong to the halo of VCC~1226 (M49), and only 18 ETGs from ACSVCS are real DSC hosts. 
Parallel to Table~\ref{gal_DSC}, we list their basic parameters in Table~\ref{galV_DSC}. \\


\section{Galaxies with DSCs}
\label{galaxy}

Since not all galaxies contain DSCs, the natural question to ask is how these DSC host galaxies are special. 

First, in our sample, the DSC hosts include both low-mass and massive ETGs. A large fraction of the massive hosts are S0 galaxies, indicating that disk environment may be important for DSCs. In addition, although some galaxies are classified as elliptical galaxies in \citet{Ferguson_89a}, most of them look like containing disks from our images. However, in both clusters, not all disk galaxies have number excess of DSC-like objects. 
Second, in both clusters, some giant elliptical galaxies are DSC hosts. Furthermore, three low-mass host galaxies, FCC~202 from Fornax, and VCC~1192 and VCC~1199 from Virgo, contain star cluster systems of their nearby massive galaxies NGC~1399 and M49 (VCC~1226) respectively, implying the existence of DSCs in the halos of massive ETGs. 
Third, the merger remnant in each galaxy cluster (FCC~21 and VCC~798) have the highest number of DSCs in their respective clusters, which indicates that galactic merger is an efficient DSC producer. 
Last but not least, six DSC host galaxies in Virgo and two in Fornax contain significant amounts of dust, showing a possible relation between DSC detection and recent star formation. Especially, the two dusty hosts in Fornax, FCC~21 and FCC~167 (NGC1380), are the galaxies with the richest DSC systems, and the third richest DSC system NGC~1340 also has wispy dust and shells. Nonetheless, not all dusty galaxies in these two galaxy clusters contain DSCs. 

\begin{figure}
\epsscale{1.2}
\plotone{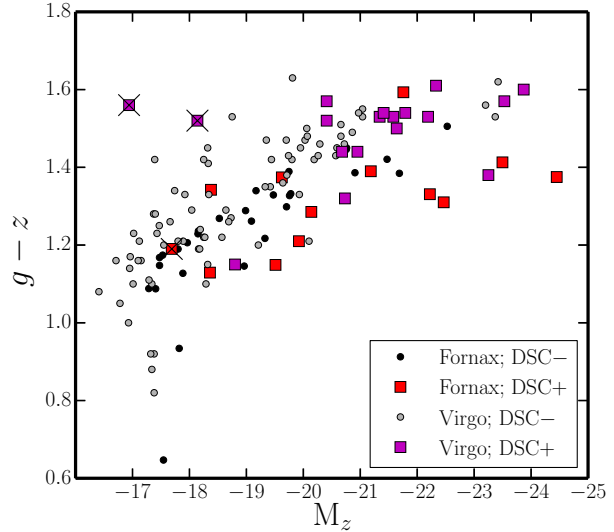}
\caption{ $g-z$ color vs. $z$-band absolute magnitude of ACSFCS (black and red) and ACSVCS (grey and magenta) galaxies. Red and magenta squares are the galaxies with DSC excess in two clusters. 
In general, the DSC host galaxies follow the general distribution well, but see the text for more detailed discussions on outliers and trends. \\
  \label{gal_CMD} }
\end{figure}

\begin{figure}
\epsscale{1.2}
\plotone{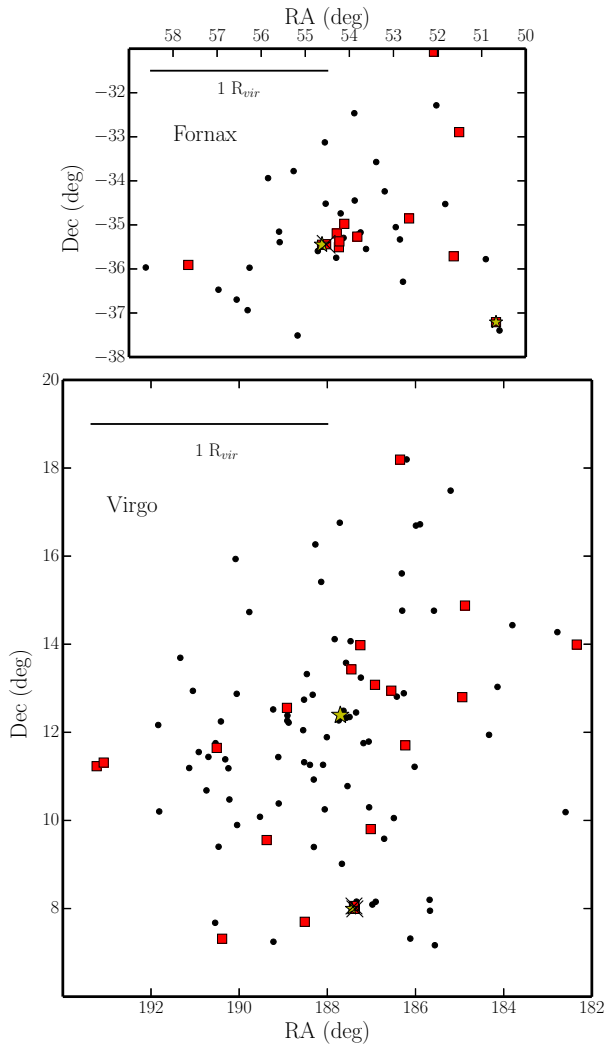}
\caption{ The location of 43 galaxies in the Fornax Cluster (\textit{top panel}) and 100 galaxies in the Virgo Cluster (\textit{bottom panel}). The big and small yellow stars are the first and second BCGs in each cluster. The galaxies with significant number of DSCs (13 in Fornax and 19 in Virgo) are marked with red squares. The three satellite galaxies FCC~202, VCC~1192 and VCC~1199 are marked with crosses. The scales of the Viral radii are displayed at top-left of two panels. \\
  \label{gal_posi} }
\end{figure}

Then we investigate whether the internal properties or external environments cause the uniqueness of DSC host galaxies. 
Figures~ \ref{gal_CMD} and \ref{gal_posi} display their positions in the galactic color-magnitude diagram and spatial distributions, but they occupy the same region of parameter space as normal galaxies. 

\begin{figure}
\epsscale{1.2}
\plotone{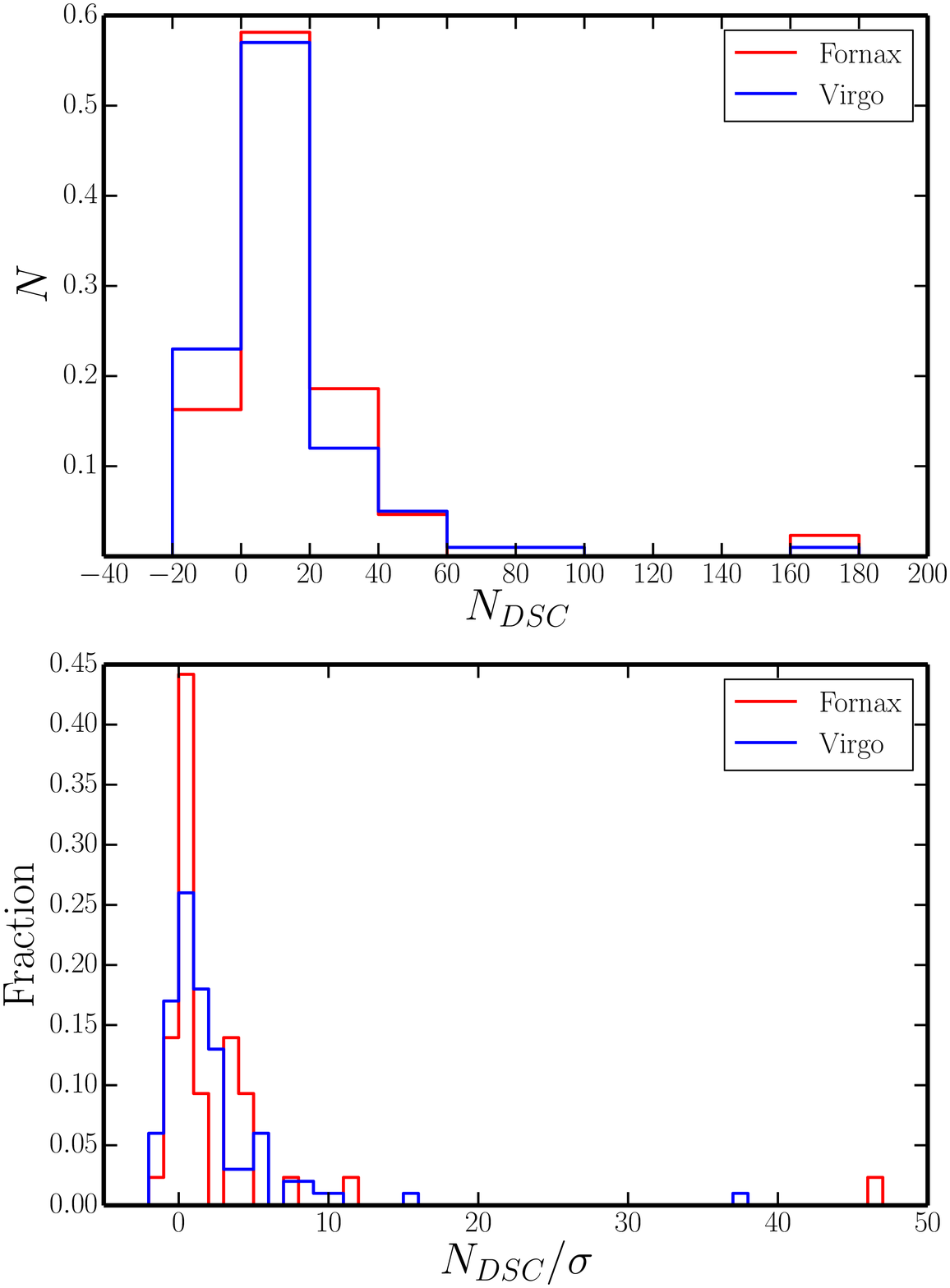}
\caption{ The normalized histograms of $N_{DSC}$ \textit{(upper panel}) and $N_{DSC}/\sigma$ (\textit{bottom panel}) of the ETGs from ACSFCS (red) and ACSVCS (blue) samples. In both panels, the K-S test of the distributions in Fornax and Virgo show high similarity, indicating that the frequency of DSC hosts is independent with the environment of their location. \\ 
\label{N_sgm} }
\end{figure}

Figure~\ref{gal_CMD} shows the $g-z$ color vs. $z$-band absolute magnitude of ACSFCS and ACSVCS galaxies. Red and magenta squares and black and grey circles indicate the DSC hosts and non-hosts in the Fornax and Virgo clusters respectively. 
The DSC systems of the three faintest hosts with crosses belong to the halos of their massive neighbors, and we only focus on the data points without crosses in this plot. 
DSC host galaxies generally follow the same broad color-magnitude distributions as others. However, there are slight differences between the two clusters. While the hosts in Virgo are mostly massive galaxies and lie on the same relation as the non-hosts, the hosts in Fornax cluster spread over a large mass range and half of them are at the blue edge of the distribution. In addition, nearly all the bluest galaxies at fixed mass in Fornax are associated with DSCs. In Virgo, only two galaxies are significantly bluer than the distribution, and one of them is a low-mass galaxy. 
One caveat is that the ACSVCS sample is not complete at low-mass ($B_T>12$), and the potential DSC hosts we missed could have special properties. 

Figure~\ref{gal_posi} shows the locations of 43 galaxies in the Fornax Cluster (\textit{top}) and 100 galaxies in the Virgo Cluster (\textit{bottom}). The scales of the Viral radii are displayed at top-left of two panels. The big and small yellow stars are the first and second BCGs in each cluster, and the galaxies with significant number of DSCs in two galaxy clusters (13 in Fornax and 19 in Virgo) are marked with red squares. The three satellite galaxies FCC~202, VCC~1192 and VCC~1199 are marked with crosses. 
These galaxies distribute evenly across the full range of cluster-centric distances in both clusters. While the DSC hosts distribution in Fornax are concentrated in the central region, unlike Virgo, it may be biased by the more centrally concentrated distribution of all galaxies in Fornax. We performed a K-S test and found that the non-similarity of the cumulative radial distributions of DSC hosts and our entire sample in the Fornax is only 0.21, with the p-value of rejecting a null hypothesis as high as 0.74. 

We also investigate the global influence from galaxy clusters. Comparing with the Virgo cluster, in which 19 or 18 (when replacing VCC~1192 and VCC~1199 by M49) out of 100 ETGs contain significant number of DSCs, the fraction of such galaxies is slightly higher in the Fornax cluster. However, this might be due to selection effects, as ACSFCS has a more complete sample than ACSVCS. In ACSVCS, 63 low-mass galaxies or S0s with evidence of recent star formation, which are possibly DSC host candidates, are missed. If we only focus on the brightest galaxies ($M_B<-18.94$) which are completed in both the Fornax and Virgo Clusters, the fractions of DSC hosts become 6 out of 9 and 13 or 14 out of 26 respectively, and the difference becomes smaller. 

The upper and bottom panels of Firgure~\ref{N_sgm} display the normalized distributions of $N_{DSC}$ and $N_{DSC}/\sigma$ respectively. The red and blue represent the distributions of the galaxies from ACSFCS and ACSVCS samples. 
$N_{DSC}$ is the number of DSCs selected from program images subtracted by the mean number of contaminants from 16 control fields, and $\sigma$ is the standard deviations of background galaxies from the 16 control fields. 
In the upper and bottom panels, the K-S test of the distributions in two galaxy clusters shows high p-value of 0.99 and 0.96 at $\alpha$ of 0.166 and 0.096, indicating high similarity of them. 
Therefore, it is evidence that the frequency of DSC hosts is independent with the environment of their location. \\


\section{DSC and GC}
\label{D_G}

In this section, we compare the properties of DSCs with GCs, and investigate how the internal galactic environment relates to DSC formation. 

\subsection{Color}
\label{color}

\begin{figure*}
\epsscale{1.25}
\plotone{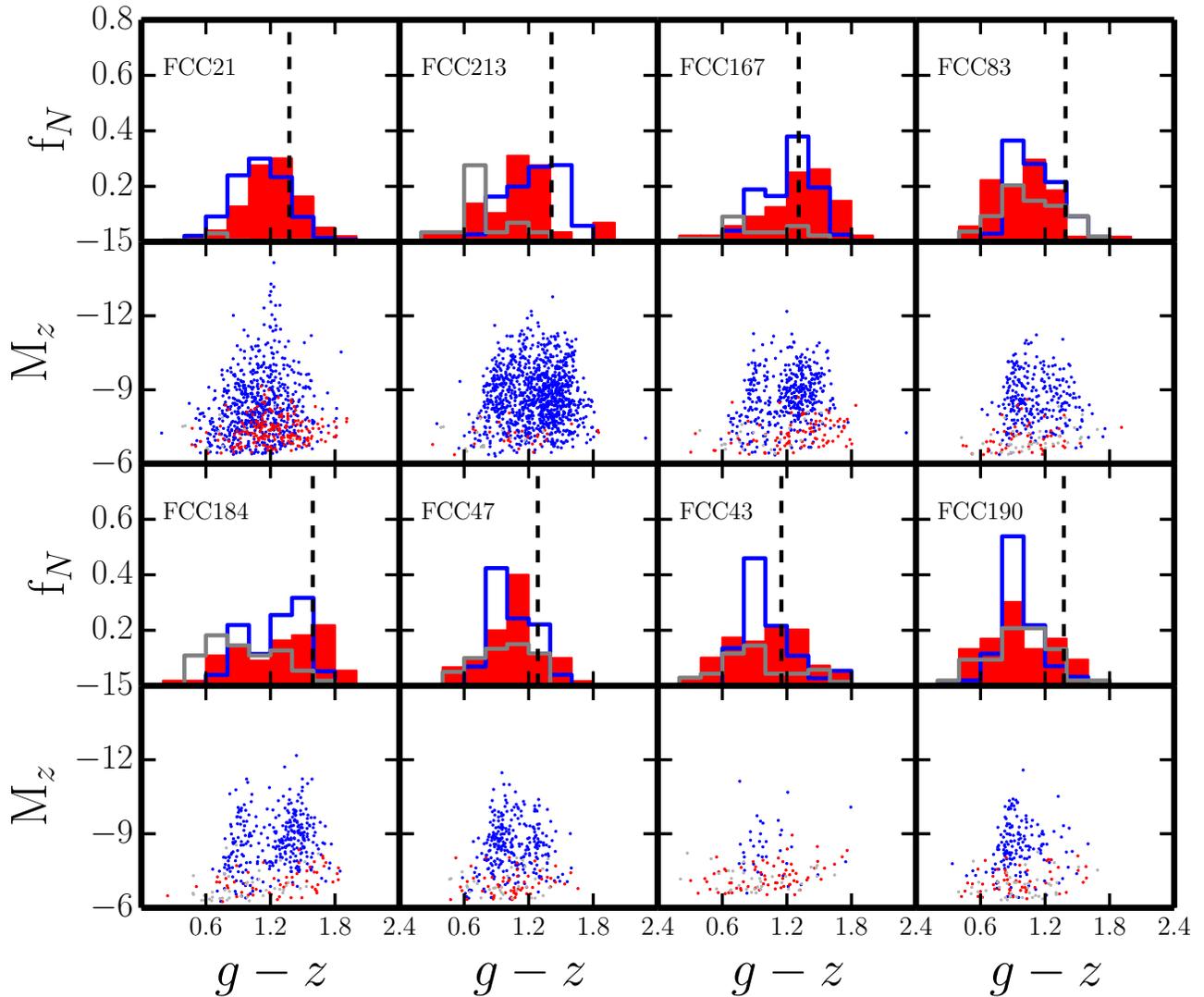}
\caption{
\textit{Odd rows}: $g-z$ color histograms (normalized) of GC (blue) and DSC (red) candidates of the 13 Fornax galaxies, as well as the entire Virgo DSC-excess galaxies. Grey histograms show the distributions of DSC-like contaminants from a random chosen control field. Dash lines represent the color of 13 host galaxies, and most of them are redder than their DSC systems. Most DSC systems in Fornax have similar or slightly redder color distributions comparing with GC's; while DSCs in Virgo are significantly redder. 
\textit{Even rows}: Below each histogram is the corresponding $g-z$ color vs. $z$-band absolute magnitude diagrams of GC (blue), DSC (red) candidates and DSC-like contaminants from a randomly chosen control field (grey). In both the Fornax and Virgo galaxies, DSCs at least follow one branch of the GC bimodality at faint ends. \\
  \label{histCMD} }
\end{figure*}
\begin{figure*}
\epsscale{1.25}
\figurenum{8}
\plotone{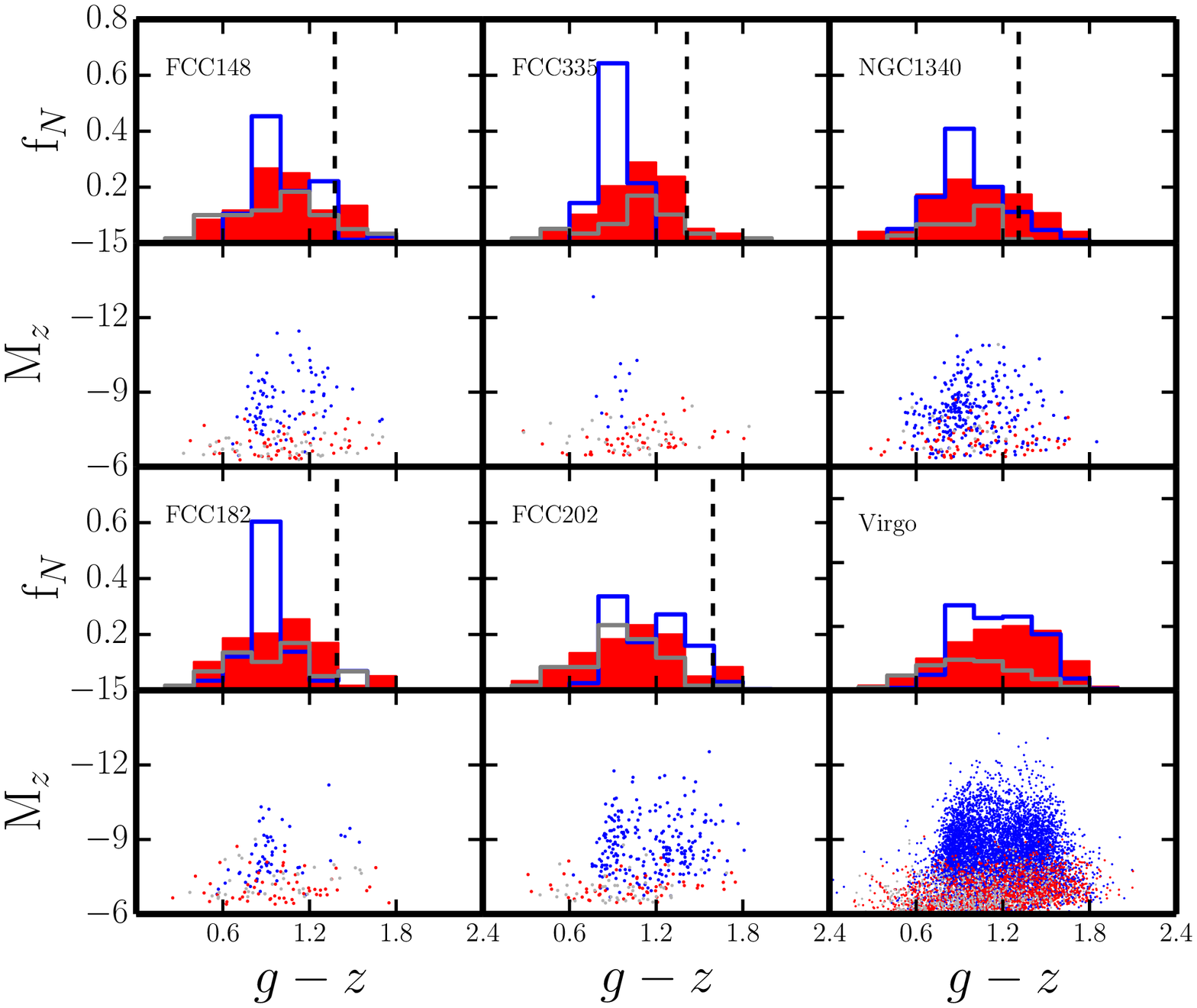}
\caption{Continued.}
\end{figure*}

The even rows of Figure~\ref{histCMD} displays the color-magnitude diagrams of GC (blue) and DSC (red) candidates in the 13 Fornax galaxies, as well as those of the entire Virgo DSC host galaxies. Grey points are DSC-like contamination from a randomly chosen control field. Above each diagram, we plot the normalized histograms of their $g-z$ color distributions in the same color coding correspondingly, and the dash lines represent the color of their host galaxies. 
Most DSC systems in Fornax have mean color similar to or slightly redder than the GC's, but bluer than the field stars of their hosts. Nonetheless, unlike Fornax, the DSCs in Virgo are significantly redder than GCs, and comparable with the field stars. 
From Figures~7 and 11 in P06, red DSCs in Virgo tend to be associated with galactic disks when dividing DSCs by $g-z=1.0$. However, when we preform the same tests on Fornax galaxies, color separation does not decouple their spatial distributions, even in FCC 335, which has a DSC system redder than GC's.

\subsection{Spatial Distribution}
\label{distribution}

\begin{figure*}
\epsscale{1.25}
\plotone{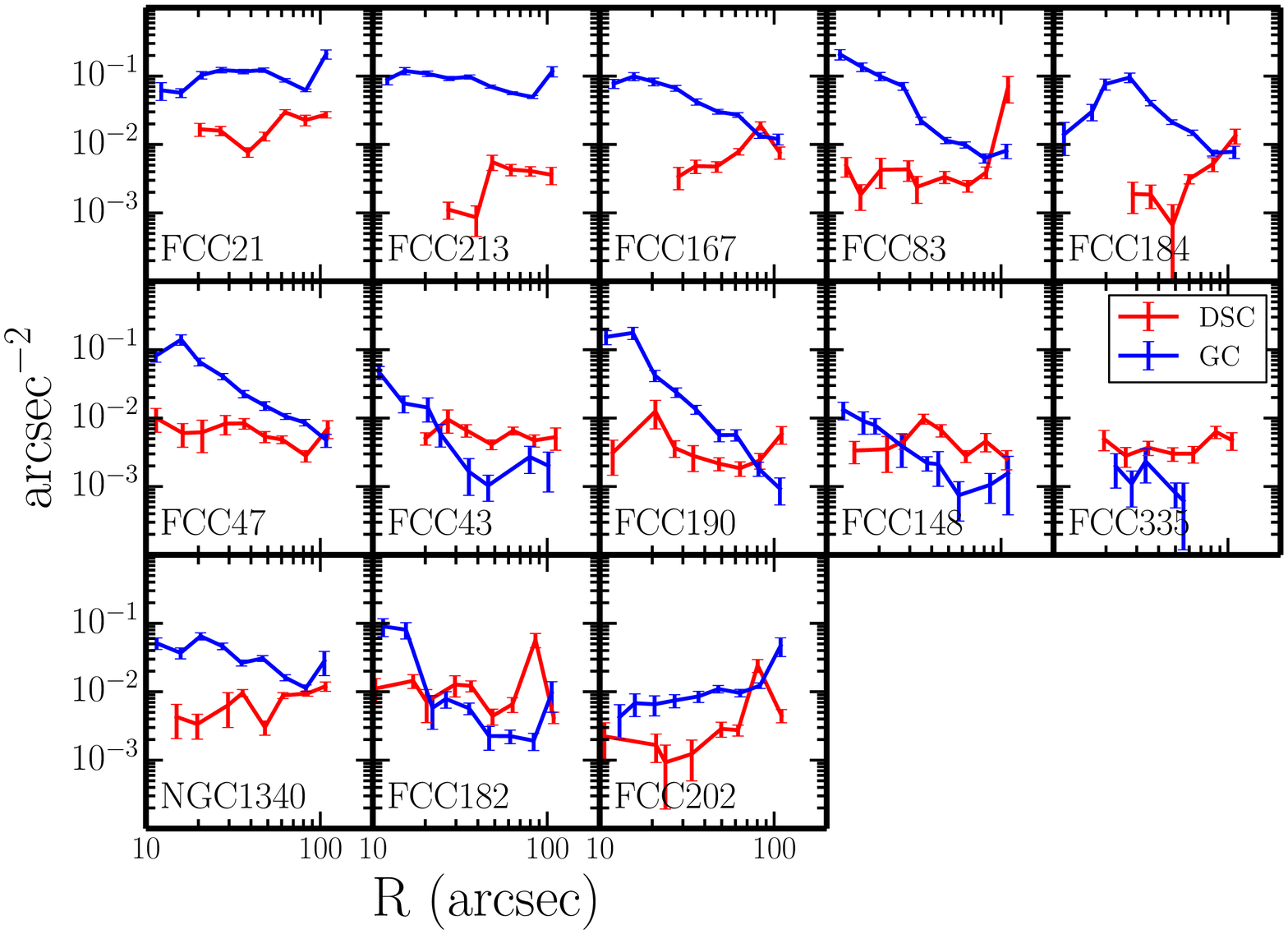}
\caption{ Radial density profiles of DSCs (red) and GCs (blue) in 13 host galaxies in Fornax. The density is calculated by third-nearest neighbor method, and given corrections on completeness. While most GC profiles are decreasing toward outer regions, all the DSC systems have flat or slightly increasing distributions. It apparently indicates that they are associated with disks, or have stronger formation/survival ability in lower density environment.  However, we cannot rule out the possibility of DSC and GC association because of the potentially high non-detection fraction in the central and brighter regions. \\
 \label{denP} }
\end{figure*}

If the formation and evolution of DSCs have connections with GCs, spatial association of these two kinds of star clusters is expected. Figure~\ref{denP} displays their radial number density profiles in 13 Fornax DSC hosts. Blue and red lines represent GCs and DSCs respectively. The density is calculated by the third-nearest neighbor method, with corrections on completeness. 
Each GC or DSC candidate is corrected by its detection probability, and we also estimate the non-detection fraction of GCs basing on their luminosity functions \citep{Villegas_10}. 

The detection probability of each source is a function of three parameters: the apparent magnitude ($m$), the size ($r_h$), and the flux of its local background ($I_b$). The detection probability is tabulated for different values of $m$, $r_h$ and $I_b$ using Monte Carlo simulations with 4,993,501 fake GCs across the full range magnitude, size, and background surface brightness. 
Specifically, for every DSC in a galaxy, we calculate the density using the third closest neighbor, corrected for detection probability. We then divide radius into 10 bins with equal logarithmic interval and calculate the mean density value in each bin. In the end, they are globally subtracted by the average density of background contaminants derived from control fields, and the data points with density lower than zero are not plotted. Because the detection probability of DSC-like objects is small and varies highly at different galaxy radius with different background brightness, and the number from control fields is not large enough to smear the random effects, we do not apply completeness correction on the control fields, and our contamination correction has no effect on the shape of radial profiles. 
For comparison, GC number density profiles are derived in the same way, except for the consideration of objects with zero detection probability. To avoid the non-detection, we do not select objects 1$\sigma$ fainter than the peak of GC luminosity function of this galaxy \citep{Villegas_10}, and divided by 0.84 for correction. A special case is FCC~21, which has a significantly fainter GC luminosity distribution. Therefore, we only select the GCs brighter than the peak of its luminosity function, and use a correction factor of 0.5. 

The density profiles of DSCs are mostly flat and possibly implying disky distributions. Some profiles are slightly increasing towards larger radii, apparently indicating their stronger formation/survival ability in lower density environment. 
The density profiles of GCs have negative gradients for most galaxies, except for FCC~21, FCC~213 and FCC~202 which have flat profiles similar to DSCs. Especially, FCC~202 is a low-mass satellite galaxy of FCC~213, and the mean densities of GCs and DSCs of FCC~202 can be regarded as the density at the outer halo of FCC~213. Thus for FCC~213, from the central region to the halo as far as FCC~202, the density of DSC remains roughly constant, while that of GC drops significantly, which is similar to most of the others. For most galaxies, the GC number densities in the central regions are higher than that of DSC. However, this is also possible to be purely due to a higher fraction of DSC non-detection in the central and brighter regions, as the difference between GC and DSC densities is smaller in fainter galaxies and at larger radii. In low-mass galaxies FCC~43, FCC~148, FCC~335, and FCC~182, the density of DSCs are comparable or even higher than that of GCs. Therefore, DSCs may be associated with GCs spatially, but we cannot detect the rise of their densities toward bright galactic centers.

\subsection{Formation Efficiency}
\label{CFE}

\begin{figure}
\epsscale{1.25}
\plotone{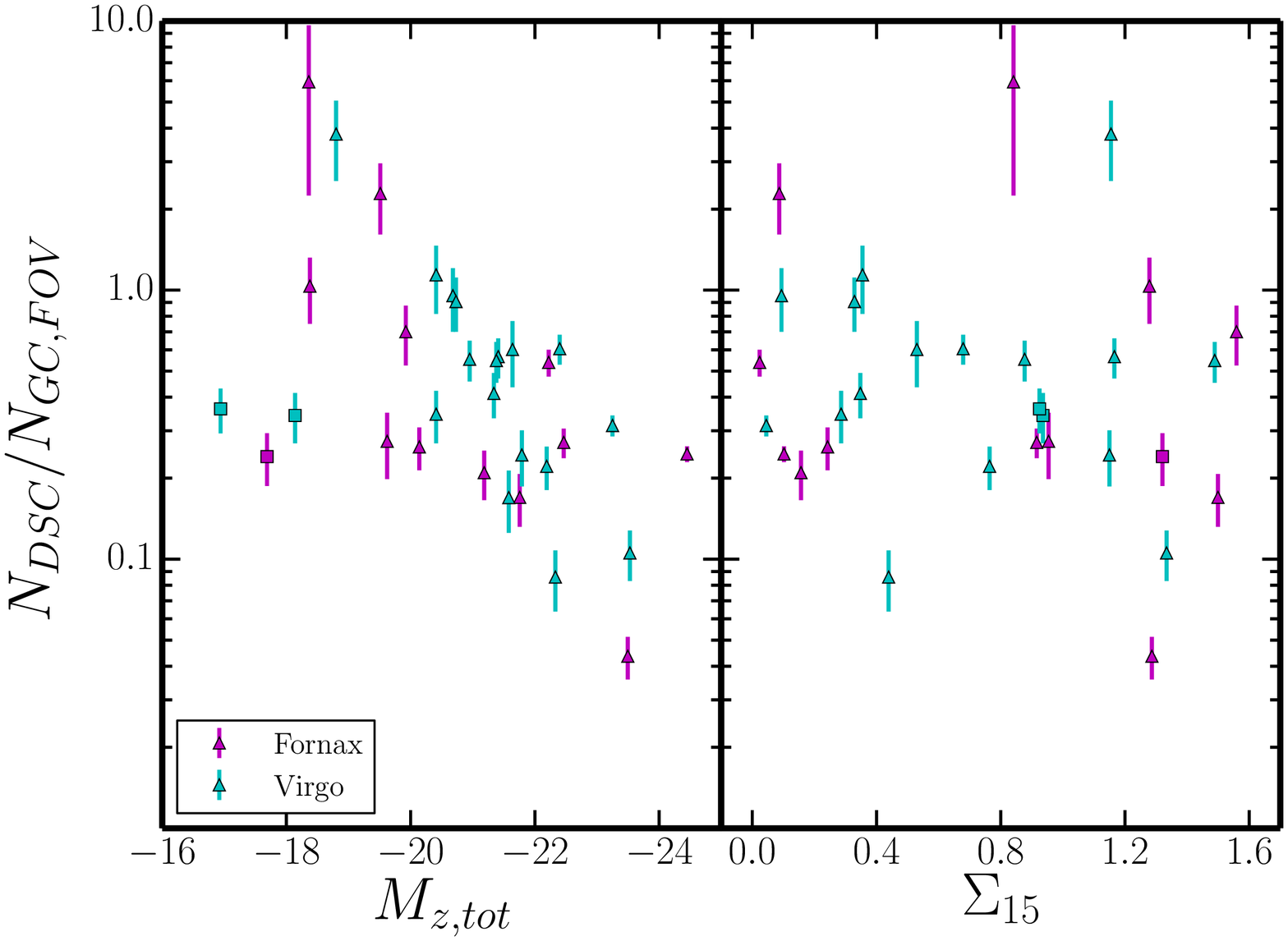}
\caption{ The comparison of the formation ability between DSCs and GCs of 32 DSC host galaxies from Fornax (magenta) and Virgo (cyan), and their relations with galactic mass and external environment. 
Three low-mass galaxies, FCC~202, VCC~1192 and VCC~1199, are marked in squares, as their star cluster systems represent the outer halos of their nearby giant ETGs NGC~1399 and M49. 
The y-axis is the number ratios between DSCs and GCs within the ACS FOV, and the numbers are corrected by detection completeness, but the non-detection fraction is not considered for DSCs. 
The left panel shows the relation with galactic $z$-band absolute magnitude, which represents their stellar mass. The ratios of low-mass galaxies are systematically higher, and the dependence on galactic luminosity is weak among massive galaxies. Specially, the ratios of three satellites have similar values to those of the intermediate mass galaxies, indicating connections between the formation and evolution of DSCs and GCs across a wide mass range of galactic environments. 
When plotting against $\Sigma_{15}$, an indicator of external environmental density in the right panel, we find no dependence on it. \\
  \label{Ngd} }
\end{figure}

Because of the potentially higher non-detection fraction in the central and brighter regions of galaxies, the flat DSC profiles shown in Figure~\ref{denP} might actually rise in the central region and follow that of GCs. To further investigate the relationship between DSCs and GCs in their formation and evolution, we compare their formation efficiency. Figure~\ref{Ngd} displays the 
number ratio between DSCs and GCs within the ACS FOV of 32 host galaxies from Fornax (magenta) and Virgo (cyan). Three squares at low-mass end represents FCC~202, VCC~1192 and VCC~1199, in which the DSC systems may belong to the nearby giant ETGs NGC~1399 and M49, and representing the properties of their outer halos. 
The numbers of DSCs and GCs are calculated using a similar method to what used for Figure~\ref{denP} and Section~\ref{distribution}. These are the sum of all objects corrected by their detection probability and background contaminants, with additional consideration for non-detection for GCs. 

The left panel of Figure~\ref{Ngd} shows the relation between the number ratio and total galactic $z$-band absolute magnitude, which represents their total stellar mass and the potential well depth. Over the full mass range, the ratio decreases as the galactic luminosity increases. 
However, this trend is mainly driven by the low-mass galaxies and the ETGs at massive end. At the low-mass end, except for the three satellite galaxies that represent the halos of massive galaxies, the ratios are systematically higher. For some objects, the ratios are even larger than unity, indicating a more efficient formation for DSCs than GCs. 
For the intermediate mass galaxies ($-23<M_z<-19.6$), the Pearson correlation coefficient is 0.35, showing a weak correlation. 
For the most massive galaxies ($M_z<-23$), because of their brighter background luminosity, the ratios of them are expected to be higher than others. 
Specially, the ratios of three satellites are similar to those of the intermediate mass galaxies, indicating similar formation efficiency at their outer halos. Therefore, there may be connections between the formation and evolution of DSCs and GCs across a wide range in mass and galactic environments. 

The right panel shows their relation with environmental density. $\Sigma_{15}$ is an indicator of environment, which is defined as the number of galaxy per square degree within a region that includes the 15 closest neighbors \citep{Guerou_15}. As shown in $\S$~\ref{galaxy}, there is no dependence with the external environment. 

\begin{figure}
\epsscale{1.25}
\plotone{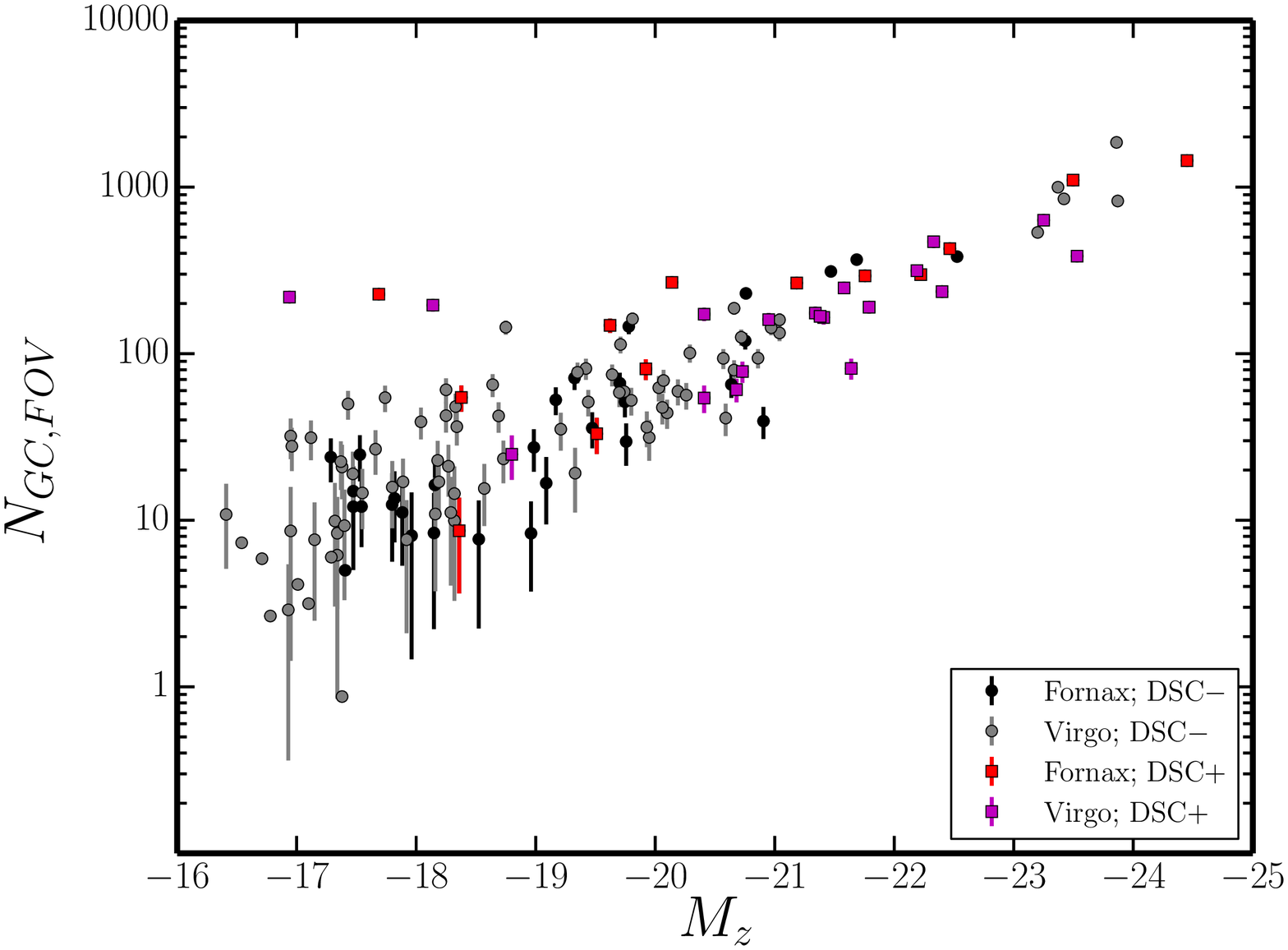}
\caption{The relation between the galactic $z$-band absolute magnitude and the number of GCs within their images. The black and grey circles represents the DSC non-hosts in Fornax and Virgo, and the red and magenta squares represents the DSC hosts in these two clusters respectively. The three outliers at low-mass with high numbers are the satellites of the nearby massive ETGs, and the GCs inside belong to their host galaxies NGC~1399 and M49. Except for these three low-mass galaxies, the GC numbers of the DSC hosts and non-hosts at similar magnitude do not show systematical offsets, implying that the formation of GCs and DSCs are independent and do not have direct effects on each other. \\
  \label{NgcMz} }
\end{figure}

Furthermore, we investigate whether the number of GCs is systematically different in DSC host galaxies. Figure~\ref{NgcMz} shows the relation between the galactic $z$-band absolute magnitude and the number of GCs within their images. The black and grey circles represents the DSC non-hosts in Fornax and Virgo, and the red and magenta squares represents the DSC hosts in these two clusters respectively. The three outliers at low-mass with high numbers are the satellites of the nearby massive ETGs, and the GCs inside belong to their host galaxies NGC~1399 and M49. Except for these three low-mass galaxies, the GC numbers of the DSC hosts and non-hosts at similar magnitude do not show systematical offsets. It implies that the formation of GCs and DSCs are independent and do not have direct effects on each other. 

Besides, the number of GCs increases with the galactic luminosity, even if only taking into account the GCs within the FOV for those massive galaxies. From the tables and Figure~\ref{findDSC}, the number of DSCs does not have large scatter among the galaxies fainter than $M_z\sim22$, and the variation of the number ratios between DSCs and GCs shown in Figure~\ref{NgcMz} are mainly driven by the number of GCs. \\


\section{Discussion: The Origin of DSCs}
\label{discussion} 

\subsection{Low-density Environment: Formation or Survival?}
\label{low-density}  

From literature, DSCs are detected in three kinds of environment: disk (spiral or S0) galaxies, low-mass galaxies, and galactic halos, all of which have relatively low density. In our sample, although some DSC host galaxies are classified as elliptical galaxies from old studies of \citet{Ferguson_89a}, they show disk-like structures in the ACS images. Especially, their DSC systems have disk-like distributions. 

One possible scenario is that the peaks of the formation radius distributions of initially bound star clusters vary with environment, with the clusters being more bound in denser environments \citep{Elmegreen_08}. During galactic evolution, less-bound star clusters with larger $r_h$ are disrupted in higher density regions \citep{Gnedin_99}, and the low-density bound DSCs are left in the moderately low density environment. 

Such a picture may explain why some low-density environments are associated with DSCs while others are not. A test for this scenario is to compare the ages of stellar disks in host galaxies with and without DSCs. If DSCs are created in all disk equally at the beginning, but we only detect the ones that have not been disrupted as time passes, then the disks containing DSCs are expected to be younger. 

\begin{figure}
\epsscale{1.2}
\plotone{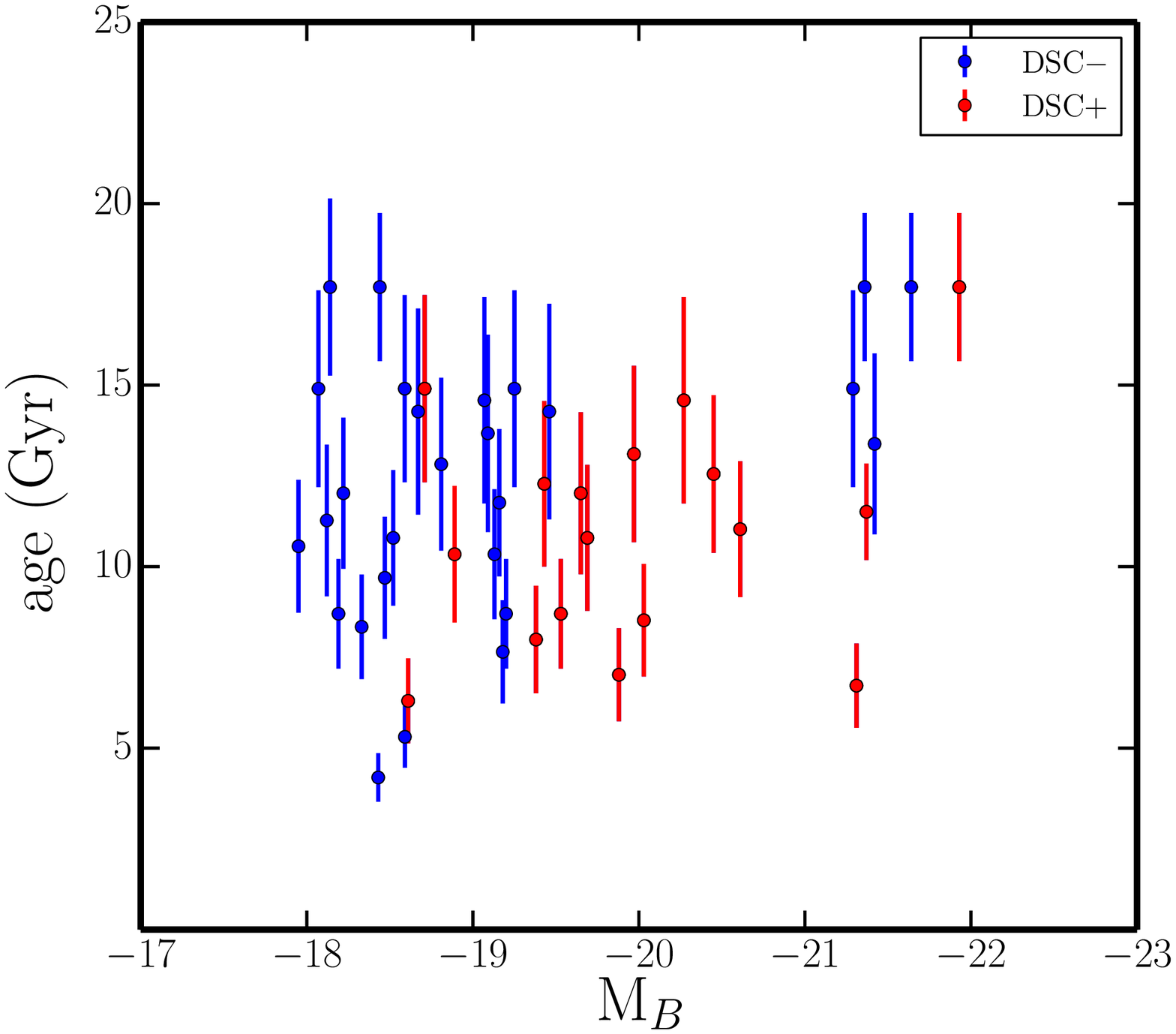}
\caption{ The mean stellar ages (y-axis) measured within 1~R$_e$ of the ETGs in the Virgo Cluster, which is a subsample of the ACSVCS. The age measurement is from \citet{McDermid_15}, and the B-band absolute magnitude is derived from \citet{Mei_07}. 
Except for two galaxies at bright end, the DSC hosts do not show significant younger ages comparing with the non-hosts at similar mass, indicating that the reasons why DSCs are only detected in a fraction of low-mass environments are from their formation instead of survivals. \\
  \label{disk_age} }
\end{figure}

Figure~\ref{disk_age} presents the ages of the massive galaxies from the ACSVCS sample that overlap with the ATLAS$^{3D}$ sample \citep{Cappellari_11} and have stellar population measurements from \citet{McDermid_15}. Red and blue circles are galaxies with or without significant DSC number excess. The ages (y-axis) are measured within 1~R$_e$, and the $B$-band absolute magnitude (x-axis) is derived from \citet{Mei_07}. 
Except for two galaxies at bright end, DSC hosts and non-hosts have similar ages at similar mass. Therefore, we suggest that the mechanisms by which the DSCs are only detected in a fraction of low-mass environments are are related to their formation instead of survival.

\subsection{Galactic Mergers and DSCs}
\label{merger} 

In our sample, two merger remnants FCC~21 and VCC~798 are both DSC hosts, indicating that a galactic merger can trigger DSC formation. M51 is another example, which is an interacting system that hosts a number of DSCs. 
At the same time, however, there are also DSC hosts containing thin disks and X-shape bulges (e.g. FCC~83, FCC~148, and VCC~2095), which are impossible to have experienced merger events. 
Therefore, DSCs may have multiple origins, either low-density environments or galactic merger events. 

Alternatively, these two environments might essentially have the same physical conditions for DSC formation. From Figure~\ref{Ngd}, FCC~21 have a similar DSC to GC number ratio to the other host galaxies, which supports this assumption. Although VCC~798 has a lower ratio, this may be due to the higher non-detection fraction. 

On the other hand, because they are brighter and have higher non-detection fraction than others, their number ratios could be substantially higher. In this case, special DSC formation mechanisms may play a role during galactic mergers.

\subsection{Other Origins}
\label{strip_expanding}

Because DSCs have relatively large sizes and diffuse light distributions, GC expansion and stripping from galaxies are two other candidates of their origin. 

In the former case, \citet{Assmann_11} tested whether a DSC similar to Scl-dE1 GC1 can form during the early evolution of a normal star cluster through gas expulsion or stellar mass loss. They found that without the embedded dark matter halos, this scenario requires the star formation efficiency of at least 0.33, which is significantly higher than what observed. 
Alternatively, tidal forces may extend GCs. However, the flat density profiles in Figure~\ref{denP} indicate no environmental dependence of DSCs inside a galaxy, which does not support this scenario. 

As for stripping, there are luminous and large star clusters that show evidence of being stripped remnants of larger systems, like UCDs in Virgo (e.g., \citealt{Zhang_15, Liu_15}). However, this mechanism is not likely for our sample, since the existence of DSCs has no preference to the outer halos of galaxies, nor the denser environment in galaxy clusters. \\


\section{Summary}
\label{conclusion}

From the images taken by ACSFCS, we find 12 out of 43 ETGs in the Fornax Cluster containing DSC-like objects more than the typical background galaxies at 3$\sigma$ level. The Virgo Cluster is the only other cluster environment with DSC detection. P06 found 12 DSC host galaxies in Virgo using the ACSVCS images of 100 ETGs, and we select out 18 hosts using the same criteria as for Fornax. In this work, we combine these two samples of 143 cluster ETGs and systematically study how the properties of DSCs relate with their host environment and GCs, in order to constrain their formation mechanisms. The main conclusions are listed as follow: 

\begin{itemize}

\item The 30 DSC hosts in our sample consist of low-mass ETGs, S0s, post-starburst merger remnants, as well as elliptical galaxies. Most elliptical galaxies contain potential disk features, except for NGC~1399, the BCGs of the Fornax Cluster. Both galaxy disks and low-mass galaxies have relatively low-density environment, indicating that DSCs can form in merger processes or low-density environments. It is possible that the physical origin of DSCs is essentially the same in these two environments, if merging places also has small tidal field. 

\item A significant fraction of massive DSC host galaxies contain dust or shell-like structures, implying that the DSC formation is related with merger and recent star formation process. 

\item Though all the DSC systems in our sample show flat galactic radial number density profiles and do not follow the distribution of GCs, the potential relations between their formation are shown in their similar color-magnitude distributions and nearly constant number ratios among the massive galaxies. The number ratios in low-mass galaxies are systematically higher, indicating a more efficient formation of DSCs in lower density environment. 

\item No evidence shows that DSC formation has any dependence on the environment of their host galaxy locations inside a galaxy cluster. 

\item In the end, why DSCs are not detected in all disky or low-mass early-type galaxies is still a puzzle. The mean ages of DSC hosts and non-hosts are similar at similar luminosities, suggesting that the reasons lie with formation history, rather than in the survival fraction. 

\end{itemize}


\acknowledgements

YL thanks for the great academic environment at KIAA. She also thanks for the helpful discussions with Long Wang on star cluster dynamics, as well as useful comments from Weijia Sun and Jinyi Shangguan. 
YL and EWP acknowledge support from the National Natural Science Foundation of China under Grant Nos. 11573002, and from the Strategic Priority Research Program, ``The Emergence of Cosmological Structures'', of the Chinese Academy of Sciences, Grant No. XDB09000105. \\


\bibliographystyle{apj}
\bibliography{ref_DSC}

\clearpage 

\end{document}